\documentclass[fleqn,10pt]{wlscirep}
\usepackage[utf8]{inputenc}
\usepackage[T1]{fontenc}
\usepackage{lineno}
\title{Training-free image style alignment for self-adapting domain shift on handheld ultrasound devices} 

\author[1,2,3]{Hongye Zeng}
\author[4]{Ke Zou}
\author[5]{Zhihao Chen}
\author[1]{Yuchong Gao}
\author[1,2,3]{Hongbo Chen}
\author[1]{Haibin Zhang}
\author[6]{Kang Zhou}
\author[7]{Meng Wang}
\author[8]{Rick Siow Mong Goh}
\author[8]{Yong Liu}
\author[9]{Chang Jiang}
\author[1,10*]{Rui Zheng}
\author[8*]{Huazhu Fu}
\affil[1]{School of Information Science and Technology, ShanghaiTech University, Shanghai, China}
\affil[2]{Shanghai Advanced Research Institute, Chinese Academy of Sciences, Shanghai, China}
\affil[3]{University of Chinese Academy of Sciences, Beijing, China}
\affil[4]{National Key Laboratory of Fundamental Science on Synthetic Vision, College of Computer Science, Sichuan University, Chengdu, China}
\affil[5]{College of Intelligence and Computing, Tianjin University, Tianjin, China}
\affil[6]{Department of Computer Science and Engineering, The Chinese University of Hong Kong, Hong Kong, China}
\affil[7]{Beth Israel Deaconess Medical Center, Harvard Medical School, Boston, USA.}
\affil[8]{Institute of High Performance Computing, Agency for Science, Technology and Research, Singapore, Republic of Singapore}
\affil[9]{Department of Orthopedics, Zhongshan Hospital, Fudan University, Shanghai, China}
\affil[10]{Shanghai Engineering Research Center of Intelligent Vision and Imaging, Shanghai, China}
\affil[*]{zhengrui@shanghaitech.edu.cn, Fu\_Huazhu@ihpc.a\-star.edu.sg}

\begin{abstract}
Handheld ultrasound devices face usage limitations due to user inexperience and cannot benefit from supervised deep learning without extensive expert annotations. Moreover, the models trained on standard ultrasound device data are constrained by training data distribution and perform poorly when directly applied to handheld device data. In this study, we propose the Training-free Image Style Alignment (TISA) framework to align the style of handheld device data to those of standard devices. The proposed TISA can directly infer handheld device images without extra training and is suited for clinical applications. We show that TISA performs better and more stably in medical detection and segmentation tasks for handheld device data. We further validate TISA as the clinical model for automatic measurements of spinal curvature and carotid intima-media thickness. The automatic measurements agree well with manual measurements made by human experts and the measurement errors remain within clinically acceptable ranges. We demonstrate the potential for TISA to facilitate automatic diagnosis on handheld ultrasound devices and expedite their eventual widespread use.
\end{abstract}
\begin{document}
\flushbottom
\maketitle
%
%
\thispagestyle{empty}

\section*{Introduction}
The development of handheld ultrasound devices has been rapid, with the devices being especially useful in rural healthcare and developing countries due to their usability and affordability in resource-constrained environments \cite{stewartTrends2020,bakerImpact2021,frijaHow2021}. Many of them are portable systems with low prices, significantly more flexible and cost-effective than standard full-sized ultrasound systems \cite{rangerPortable2023}. A common barrier to the widespread adoption of handheld devices is inexperience with the technology by healthcare providers due to their limited knowledge and experience with ultrasound \cite{abrokwaTask2022,ginsburgSurvey2023}. Artificial intelligence (AI) technologies can help augment the capabilities of these devices in disease assessment and diagnosis to mitigate deficiencies in user knowledge \cite{shenDeep2017,liuDeep2019}.

There are two main obstacles to applying current AI models for handheld devices: (1) existing supervised learning models require extensive high-quality expert annotations \cite{litjensSurvey2017a,estevaDeep2021} that are difficult and time-consuming to acquire; (2) the models trained on standard devices are constrained by training data distribution and usually experience a considerable performance decline when confronted with data collected from different imaging systems \cite{redkoAdvances2019}. For example, the detection network trained on spine images from a standard device would face a problem with images from the handheld device, leading to failures in automatic spinal curvature measurement. The handheld device has the additional drawback of reduced imaging quality due to hardware limitations (compared with high-quality images from standard devices), which is primarily responsible for the domain-shift problem.
\begin{figure}[ht!]
\centering
\includegraphics[width=0.84\linewidth]{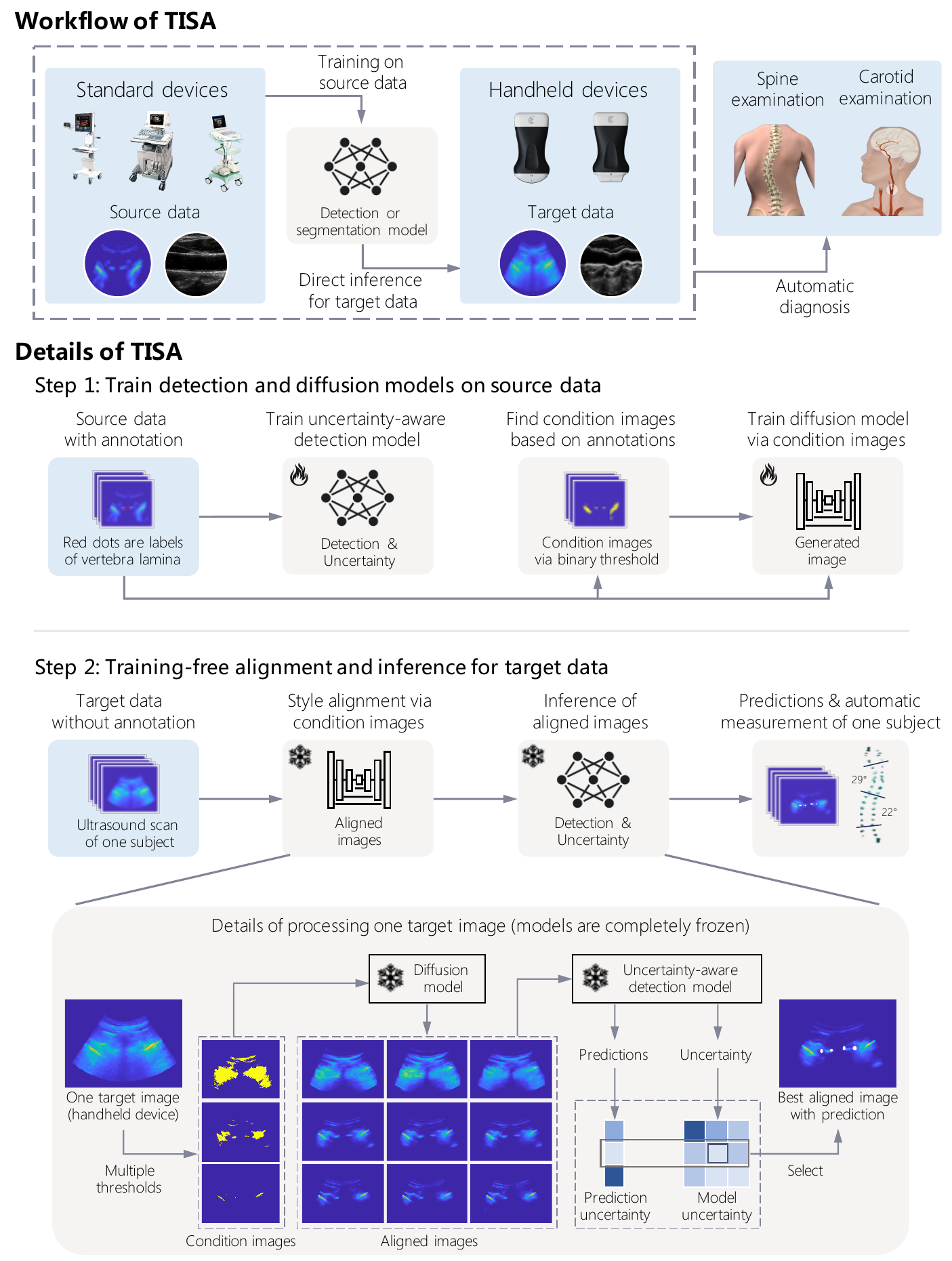}
\caption{\textbf{Training-free image style alignment (TISA) for ultrasound image analysis on handheld devices.} We trained models on source data and directly applied them to target data through TISA. The diffusion model aligned the target image style to resemble the source data while ensuring structural integrity. The uncertainty-aware model predicted all aligned images and selected the best one by using uncertainty. We collected source data from three standard devices and target data from two handheld devices through spinal and carotid examinations. We conducted the evaluations on detection, segmentation, and automatic measurements.}
\label{fig:method}
\end{figure}

Domain adaptation is one solution for this domain-shift problem, which adapts models for use from standard to handheld devices. Examples of this solution include unsupervised domain adaptation (UDA) \cite{kanakasabapathyAdaptive2021,zhaoLEUDA2023,shinSDCUDA2023a,roschewitzAutomatic2023,guoHandling2023,xieDirichletbased2022}, source-free domain adaptation (SFDA) \cite{yuComprehensive2023,yangSource2022,wuUPLSFDA2023,liModel2020a} and test-time domain adaptation (TTDA) \cite{wangTent2021,niuEfficient2022,niuStable2023,wangContinual2022}. However, deploying these approaches for handheld devices is unrealistic due to the need for model updates. One problem is that handheld devices typically contain small datasets without annotations, making it difficult to gather enough data for multi-epoch updates. Another problem arises from the significant intra-variation in handheld device data arising from flexible usage environments and inexperienced operators. Such a scenario necessitates continuous model updates to accommodate these variations, further incurring significant computational costs and leading to unstable performances.

In this study, we propose a training-free image style alignment approach for handheld ultrasound devices. The method, Training-free Image Style Alignment (TISA), can align images captured with handheld devices to fit the model trained on standard device data. In contrast to previous approaches, TISA does not require updates to the model and is better suited for use in clinical handheld devices. To demonstrate the success of our method, we evaluated the performance of TISA in medical detection and segmentation tasks involving target spine and carotid images collected from different handheld devices. We further validate TISA as a clinical model in spinal curvature measurement and carotid intima-media thickness measurement. We also show that TISA outperforms current methods in two tasks, allowing handheld devices to be used for automatic diagnoses that remain consistent with results obtained traditionally by experts. We believe that TISA holds the potential to enhance healthcare providers’ capabilities and remove barriers to the widespread use of handheld devices.

\section*{Results}
To evaluate TISA’s performance across different devices, we conducted spine and carotid examinations using handheld devices, focusing on detection, segmentation, and automatic measurements (Figure \ref{fig:method}). The detection and segmentation were conducted with TISA, which comprised of two steps: (1) training the diffusion model and uncertainty-aware model on the source data, and (2) aligning the target images to a source-like style and filtering predictions of the aligned images using uncertainty.
\\

\noindent \textbf{Data construction and study design}\\
We collected source and target data through ultrasound spinal examinations, and conducted evaluations using target data for vertebral structure detection and spinal curvature measurement. The source data was gathered from the hospital in Canada using a standard ultrasound device SonixTABLET and comprised 2163 annotated and 32,085 unannotated images. The target data was collected from the rehabilitation center in China using a handheld ultrasound device Clarius C3 with a convex array probe. It consisted of 106 adolescent ultrasound scans including 79,790 transverse slices (15 males and 91 females; age: $8\sim18$ years ($13.8\pm2.3$); height: $122.9\sim180$ cm ($159.4\pm11.5$), weight: $21\sim64$ kg ($46.1\pm9.9$), major curve: $5^{\circ}\sim55^{\circ}$). We used all source images to train the diffusion model and the annotated source images to train a robust uncertainty-aware detection model. This detection model was also utilized by other adaptation methods. We sampled and annotated 1479 slices from the target data to evaluate detection results, while all scans and slices were used to validate automatic spinal curvature measurement. The spinal curvature can be used to assess adolescent idiopathic scoliosis (AIS), which is a 3D deformity of the spine featuring lateral deviation and axial vertebral rotation \cite{weinsteinHealth2003, weinsteinAdolescent2008,asherAdolescent2006}. The ultrasound imaging technique has been clinically validated for assessing AIS on standard ultrasound devices, serving as a radiation-free alternative to the use of the Cobb angle measurements on radiographs \cite{zhengIntra2015,chenReliability2013, ungiAutomatic2020, zengAutomatic2021}. 

We also evaluated TISA through the carotid intima-media segmentation and thickness measurement on carotid target data. Carotid intima-media thickness (CIMT) is a commonly used marker for atherosclerosis risk assessment and is often computed on carotid ultrasound images by delineating the intima-media complex \cite{meiburgerCarotid2022, liAutomatic2023}. We used the carotid ultrasound boundary study (CUBS), a public carotid ultrasound dataset from standard ultrasound devices (Philips HDI 5000 in Cyprus and Esaote MyLab25 in Italy) as the source data. The data includes 1088 participants and 2176 total images acquired from both sides of the neck \cite{meiburgerCarotid2021}. We trained the diffusion and segmentation models using the source data. We collected target data from a hospital in China using a handheld ultrasound device (Clarius L7 with a linear array probe) on 59 participants and obtained 77 images for evaluation. We provided details of data construction and evaluations in supplemental Figure S1.
\\
\begin{figure}[ht!]
    \centering
    \includegraphics[width=1\linewidth]{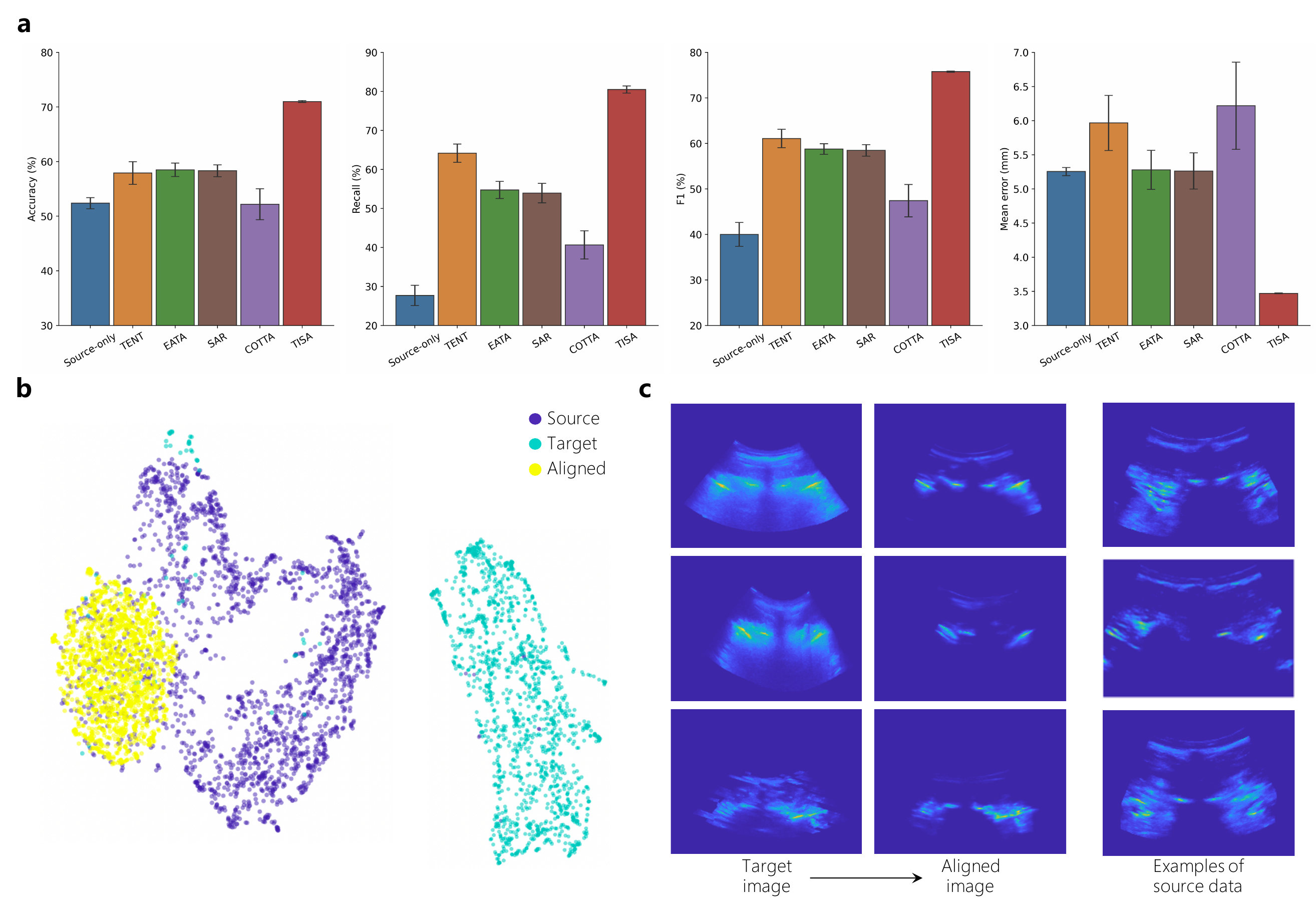}
    \caption{\textbf{Comparison of domain adaptation methods for vertebral structure detection on the handheld device data.} \textbf{a,} The accuracy, recall, F1 score, and mean error of our method compared to other adaptation methods when tested on the target data (n=1469). Each result was an average of five trials with different random initializations and error bars depicting the standard error of the mean (SEM). \textbf{b,} UMAP plots illustrating source, target, and aligned target data in the feature space. TISA effectively eliminated domain shifts between source and target data. \textbf{c,} Examples of target data and the corresponding aligned images. TISA aligned images with the source-like style while preserving structural details.}
    \label{fig:spine_2d}
\end{figure}
\begin{figure}[!ht]
    \centering
    \includegraphics[width=0.96\linewidth]{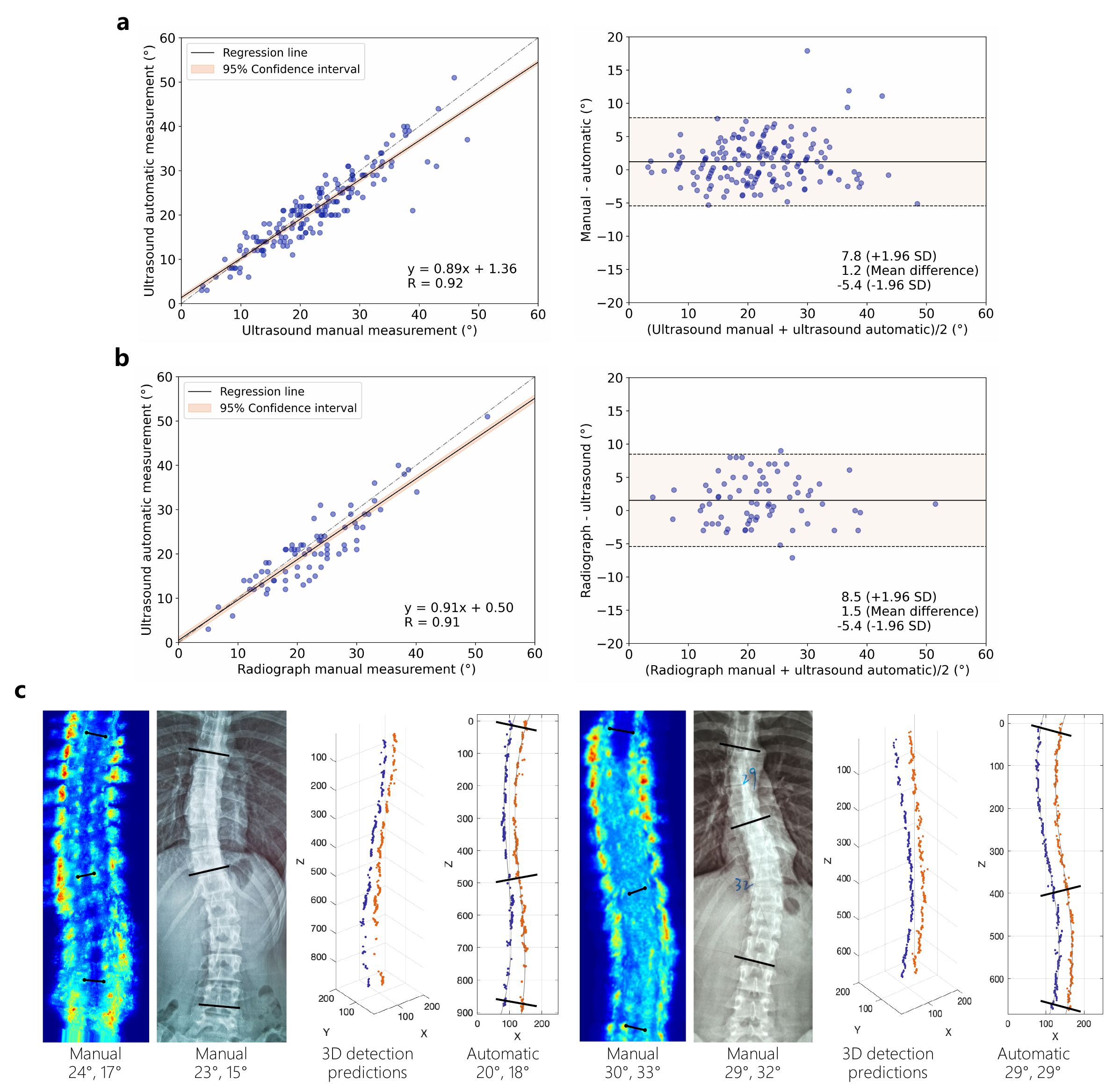}
    \caption{\textbf{Assessment of TISA in automatic spinal curvature measurement on clinical data collected from the handheld device. a,} Analysis between ultrasound manual and automatic measurements (106 subjects, 79,790 transverse slices, n = 165). The manual measurements are acquired onsite by skilled experts. \textbf{b,} Analysis between ultrasound automatic and radiographic manual measurements (49 subjects, n = 73). The radiographic measurements have been implemented at various other hospitals as the gold standard diagnostic results. We exclude cases that are too outdated or involve subjects who have received treatment. \textbf{c,} Visualization of two measurement methods on mild ($10^{\circ}\sim25^{\circ}$) and moderate ( $25^{\circ}\sim40^{\circ}$) spinal curvatures.}
    \label{fig:spine_3d}
\end{figure}

\noindent \textbf{Performance of vertebral structure detection}\\
TISA achieved an accuracy of 70.99\% (95\% confidence interval (CI) = 70.43-71.55\%), a recall of 80.49\% (95\% CI = 77.66-83.32), an F1 score of 75.75\% (95\% CI = 75.30-76.20), and mean error (ME) of 3.47mm (95\% CI = 3.45-3.49mm) in the detection evaluation of target data (Figure \ref{fig:spine_2d}a). Compared to the source-only method, which uses the detection model trained on the source data without any adaptation, TISA achieved statistically significant improvements (p < 0.001, by a two-sided paired t-test) of 18.61\% in accuracy, 52.80\% in recall, 35.76\% in f1, -1.78mm in ME, respectively. TISA can effectively eliminate domain shifts between source and target data by aligning image style while preserving structural details (Figure \ref{fig:spine_2d}b, \ref{fig:spine_2d}c). We also illustrate the variations in aligned images under different condition images in supplemental Figure S2.
\begin{figure}[ht!]
    \centering
    \includegraphics[width=0.86\linewidth]{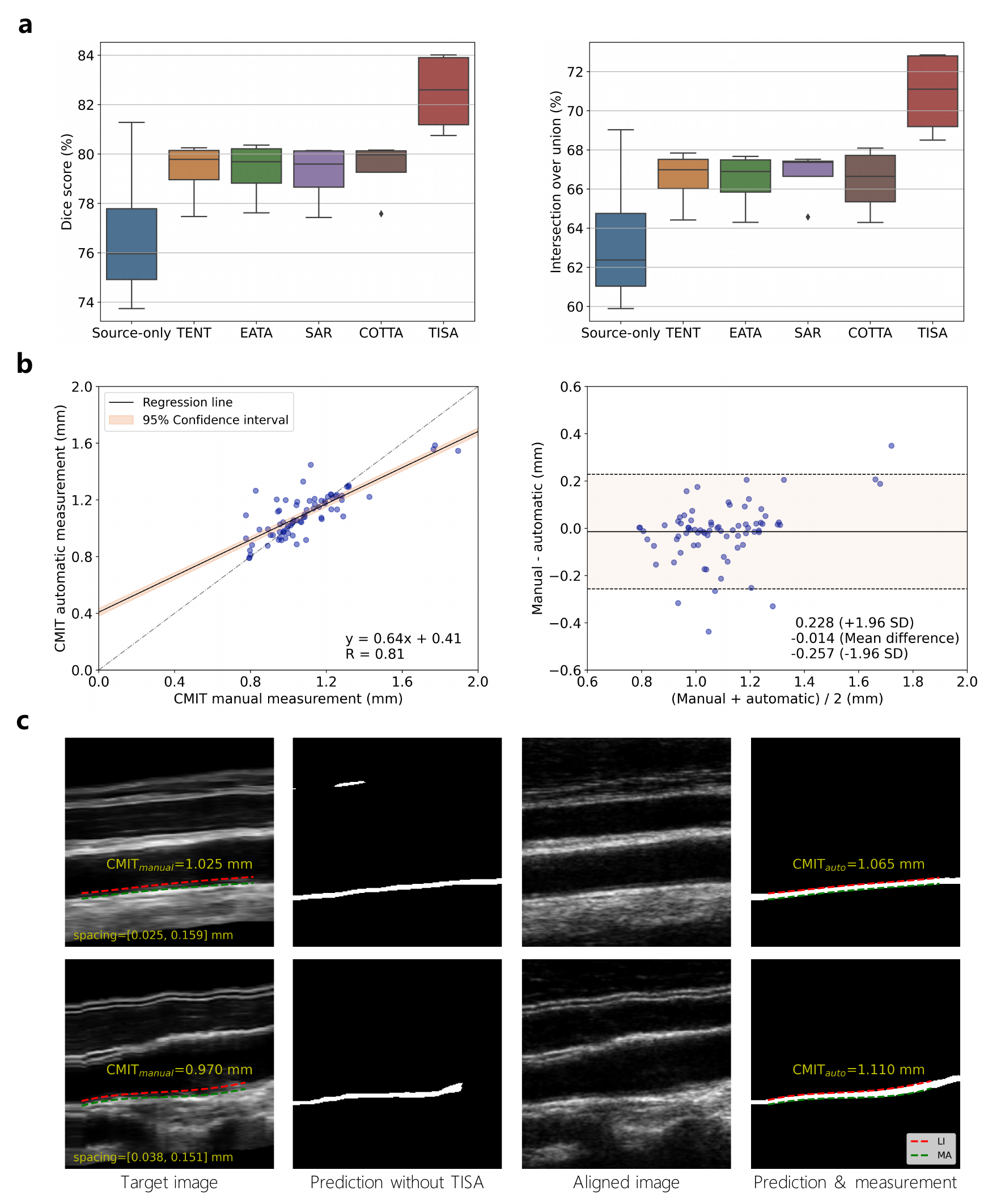}
    \caption{\textbf{Performance of TISA in carotid intima-media segmentation and thickness measurement using handheld device data. a,} Boxplot comparison of Dice score and IoU between all adaptation methods over multiple runs. \textbf{b,} Correlation and Bland–Altman analysis between manual and automatic measurement (n = 76). \textbf{c,} Aligned images with their segmentation results and automatic thickness measurements. The thickness plotted on the target image was manually measured by tracing the contours of the lumen-intima (LI, red line) and media-adventitia (MA, green line) anatomical interfaces.}
    \label{fig:carotid}
\end{figure}

Furthermore, we benchmarked our method against leading domain adaptation methods (TENT\cite{wangTent2021}, EATA\cite{niuEfficient2022}, SAR\cite{niuStable2023}, COTTA\cite{wangContinual2022}) on the target dataset, all adapting an identical trained model derived from the source data. Our method consistently outperformed these other approaches in all metrics, showing an increase of 12.5\% in accuracy, 16.35\% in recall, 14.68\% in F1 score, and a decrease of 1.79 mm in ME compared to the best in each category. Statistical analyses also confirmed that the differences are significant across all comparisons, with p-values less than 0.01. Moreover, we observed that TISA was more stable compared to other approaches. TISA achieved lower standard errors of the mean (SEMs) in accuracy (0.18\%), recall (0.91\%), and F1 score (0.14\%), as well as showing a lower ME (0.01 mm) compared to the SEMs of TENT (2.07\%, 2.35\%, 2.03\%, 0.4 mm respectively), EATA (1.23\%, 2.21\%, 1.17\%, 0.29 mm respectively), SAR (1.13\%, 2.5\%, 1.27\%, 0.27 mm respectively), and COTTA (2.82\%, 3.62\%, 3.55\%, 0.64 mm respectively).
\\

\noindent \textbf{Performance of automatic spinal curvature measurement.} \\
Once an ultrasound scan was acquired, skilled experts manually measured the spinal curvature onsite. In addition, 85 subjects had undergone radiographic examinations and the Cobb angles were measured as part of the diagnoses conducted in hospitals. We excluded cases where the interval between the radiograph and ultrasound scans exceeds 90 days or where subjects had received brace treatments in the interim, as these factors can cause significant discrepancies in spinal curvature. Ultimately, the Cobb angles from 49 subjects were available for comparison between radiographic and ultrasound measurements.

We detected vertebral structures using our TISA framework for each ultrasound scan and automatically measured these detection results (Figure \ref{fig:method}b step 2). Supplemental Figure S3 shows that vertebral structures are seldom detected without employing TISA. Compared to the manual ultrasound measurements (Figure \ref{fig:spine_3d}a), our automatic method showed a mean absolute difference of $2.6^{\circ}\pm2.3^{\circ}$, a high correlation coefficient of 0.92 (p < 0.0001), and $R^2$ of 0.85. This was well within the range of typical measurement variation ($5^{\circ}$) between different methods usually ascribed to inter-observer variation \cite{cobbOutline1948}. A Bland–Altman analysis indicated a mean bias of $1.2^{\circ}$ (standard deviation (s.d.) = $3.4^{\circ}$) with limits ranging from $-5.4^{\circ}$ to $7.8^{\circ}$ (n = 165). Our automatic method showed no systematic biases but exhibited proportional biases. When compared with manual radiographic measurements (Figure \ref{fig:spine_3d}b), the automatic measurements also agreed strongly with a high correlation coefficient of 0.91 (p < 0.001) and $R^2$ of 0.82. The mean absolute difference was $3.1^{\circ}\pm2.3^{\circ}$. The Bland–Altman test showed a mean bias of $1.5^{\circ}$ (s.d. = $3.5^{\circ}$) with limits ranging from $-8.5^{\circ}$ to $5.4^{\circ}$ (n = 73). The measurements with the three methods on subjects with two different degrees of spinal curvature exhibited high consistency (Figure \ref{fig:spine_3d}c).
\\

\noindent \textbf{Performance of carotid intima-media segmentation and thickness measurement}\\
We applied TISA to the carotid intima-media segmentation and thickness measurement (see details in supplemental Figure S4). Our approach achieved the best Dice score of 82.49\% (95\% CI = 79.80-85.18\%) and the best intersection over union (IoU) of 70.89\% (95\% CI = 67.29-74.49\%) (Figure \ref{fig:carotid}a). The boxplot showed that TISA had the highest median, indicating that it typically performs better over multiple runs. The source-only method exhibited instability across different runs, with its Dice score ranging from 73.74\% to 81.28\%. This variability led to performance variations in TISA, as it utilized the model trained on source data without any updates. A strong positive linear correlation between automatic and manual CMIT measurements was presented with a correlation coefficient of $0.81$ (p<0.0001) and $R^2$ of 0.65, and the Bland–Altman test displayed a mean bias of 0.014mm (s.d. = 0.123mm) with limits ranging from -0.257mm to 0.228mm (n = 76). Two measurements generally remained consistent despite slight deviations. Figure \ref{fig:carotid}c showed aligned images and their segmentation predictions and automatic measurements. 

\section*{Discussion}
The handheld device data usually experiences a considerable performance decline when directly using models trained on standard device data due to the domain-shift problem. We have proposed Training-free Image Style Alignment (TISA) to adapt the domain shift between standard and handheld devices. Our experiments showed that, without any updates to models during inference, TISA performed better than other domain adaptation methods and was comparable to the performance of experts when used for automatic diagnosis as clinical models. Specifically, TISA achieved an F1 +16.68\% and a Dice score +3.07\% above that of the highest-performing domain adaptation method in vertebral structures detection and carotid intima-media segmentation (Figure \ref{fig:spine_2d} and \ref{fig:carotid}). We also showed that the performance of the diagnosis using TISA is comparable to that of experts, as there are high positive correlations and no statistically significant differences between automatic measurements of TISA and manual measurements of experts over spinal curvature and CMIT measurements (Figure \ref{fig:spine_3d} and \ref{fig:carotid}).
 
Previous efforts for the domain-shift problem have shown meaningful improvements in medical image analysis, but cannot be deployed for handheld ultrasound devices which are used in flexible and resource-constrained clinical environments. For instance, UDAs require that both the source and target domain data should be accessed during adaptation \cite{zhaoLEUDA2023,shinSDCUDA2023a}, which is unrealistic in clinical applications. The raw source data is generally unavailable due to data privacy concerns and it is impractical to keep the complete source dataset for handheld devices for training. SFDAs have solved the above issues but necessitate a substantial target dataset for multi-epoch training \cite{yuComprehensive2023}. With handheld devices being a relatively new technology, there is a significant time requirement for the accumulation of extensive data, as evidenced by our target carotid dataset that contains only 77 images. In contrast, our TISA is able to process one target image without reutilization of source data. TISA opens promising avenues for AI applications on handheld ultrasound devices, where small and isolated datasets are common. 

To be more suitable for clinical applications, the latest TTDAs \cite{wangTent2021,niuEfficient2022,niuStable2023,wangContinual2022} only need one mini-batch of target data to update the model trained on the source data. TISA has two advantages over TTDAs: (1) TISA performs better and more stably. For instance, in the detection task, TISA showed the least SEMs in accuracy (0.18\%) and F1 score (0.14\%) over different runs. TTDAs necessitate continuous updates to the model with mini-batch and suffer influences of small batch size, imbalanced data, and randomness. In contrast, TISA avoids these influences by processing one image target independently. (2) TISA incurs lower computational costs as it does not require continuous model updates, making it more conducive for handheld device deployment. Both diffusion and uncertainty-aware models are trained using the source data and are completely frozen during alignment and inference. Condition images used for style alignment also can be easily obtained through non-learning methods.

Our approach demonstrates the potential for automatic diagnoses with handheld ultrasound devices. We automatically derived spinal curvature measurements from 3D ultrasound spine images using the results of vertebral structure detection obtained through TISA. The automatic measurements agreed closely with manual measurements taken from ultrasound images and radiographs, and all average differences between automatic and manual measurements were in the range of clinical acceptance error for scoliosis assessment (Figure \ref{fig:spine_3d}). Although there were larger discrepancies in cases involving severe curvature ($>40^{\circ}$), these could be attributed to the relatively low number of subjects with severe curvature in our dataset from the rehabilitation center. Patients with severe spinal curvature are less common, and they typically do not visit rehabilitation centers, as surgery may be a more effective treatment option for them. The correlation between automatic ultrasound and manual radiographic measurements was slightly poorer and with a higher error, as ultrasound and radiography are two different imaging modalities. Crucially, the time range between the two scans in our dataset is ninety days, during which there can be changes in the subjects’ condition. Posture during the scans can also impact the results. The subjects stand in a positioning chariot for radiography, whereas they are freestanding during the ultrasound examination. Therefore, the posture of the subjects may differ between the two examinations, causing the spinal curvature measurement from ultrasound images and radiographs to diverge. We implemented automatic CMIT measurement based on segmentation results from TISA, which agreed with the manual measurements with a high degree of correlation (Figure \ref{fig:carotid}). A few large differences between the two sets of measurements can be attributed to the large spacing of pixels. The CMIT ranges from 0.776 mm to 1.895 mm in our dataset, and the vertical spacing is approximately 0.15 mm. A single-pixel difference between the predicted mask and annotation may result in a large measurement difference. Although we computed the average thickness of the mask as CMIT, it is still hard to eliminate this effect. A similar phenomenon also has been observed in an identical comparison of standard device images, which displays a bias as high as 0.255 0.230 between two manual measurements from different raters \cite{meiburgerCarotid2021, meiburgerCarotid2022}. Overall, the measurement errors remained within clinically acceptable ranges, and both two automatic measurements were largely consistent with manual measurements made by human experts.

Our study suffers from a few limitations. Firstly, the diffusion model has a slow inference speed and our need for the model for each adaptation slows down this process considerably. Recent efforts on deploying diffusion models in mobile devices \cite{yanDiffusion2023, zhaoMobileDiffusion2023} could be adapted and integrated into our work. For instance, MobileDiffusion \cite{zhaoMobileDiffusion2023} can achieve remarkable sub-second image generation on mobile devices for large-scale diffusion models by adjusting the model structure and optimizing the sampling process, displaying significant potential for enhancing our approach. Secondly, the automatic measurements still need to be validated on more clinical data, especially the CMIT measurements. Lastly, to expand the application scope of handheld devices, TISA needs to be applied to ultrasound images of a wider range of body parts for automatic measurements in diagnosing more diseases.

In conclusion, we developed TISA to allow handheld device data to be applied to models trained on standard device data. TISA is designed for use in clinical scenarios and can function without annotations from the target data, reutilization of the source data, and updates to the model. Compared to existing techniques, our approach achieved better and more stable performance in medical detection and segmentation tasks. We applied TISA to take automatic measurements using handheld devices, which matched expert-level performance on spinal curvature and CMIT measurements. Our approach showed the potential for TISA to enhance users’ capabilities in disease assessment and to broaden the usage and healthcare service coverage of handheld devices.

\section*{Methods}
Our study method consists of two steps as shown in Figure \ref{fig:method}. Firstly, we trained the diffusion model and uncertainty-aware model using the source data with annotations. Secondly, we aligned one target image to the source-like style while preserving its original spatial structure using the diffusion model. All aligned images of this target image were predicted and filtered using the uncertainty-aware model. Both two models were completely frozen during the inference of target data. We provide details on model training and train-free style alignment in the following.
\\

\noindent \textbf{Diffusion model architecture.} 
The conditional diffusion model aligns the target image to match the style of the source domain while preserving the original spatial structure. Our diffusion model is fundamentally a conditional denoising diffusion probabilistic model \cite{hoDenoising2020} (DDPM) and used the design of ControlNet \cite{zhangAdding2023} to control the generation of the DDPM. It consisted of two main components: a locked U-Net and a trainable copy of its decoder and middle block. These two components were connected in the decoder using  "zero convolution" layers —  $1\times1$ convolution layers with weights and biases initialized to zeros, and these layers were disconnected to improve the training speed. We employed the U-Net as the base network, which comprised of 6 blocks in the encoder, a middle block, and 6 blocks in the skip-connected decoder. Each block consisted of ResNet and downsample layers, and the multi-head attention layer was added to low-resolution blocks.
\\

\noindent \textbf{Train diffusion model using source data.}
The initial step involved training an unconditional DDPM for source image synthesis. Given a source image $x_s$, the forward process in the diffusion algorithm progressively adds random noise to the image $t$ times and produces the noisy image $x_s^t$. In the reverse process, the U-Net is capable of estimating noise added at the $t^{th}$ step and can be optimized by minimizing the distance between the real and the estimated noises. 
Once trained on a large volume of source data, the DDPM can generate source-like images by progressive denoising an input Gaussian noise, but currently generated images are completely random without any spatial control. Following the ControlNet strategy, we locked the U-Net of trained DDPM to preserve the generative capability learned from a large dataset and duplicate a trainable copy to learn spatial control using paired images and conditions.
The learning objective of the entire generative model $\theta$ is defined as
\begin{equation}
L_g=\mathbb{E}_{x_s,t,c,\epsilon\sim\mathcal{N}(0,1)}\Big[\left\|\epsilon-\epsilon_\theta\left(x_s^t,t,c\right)\right\|\Big]
\end{equation}
where $c$ is condition images of source data $x_s$ and is produced based on annotation. We first trained the U-Net of DDPM for 400 epochs and then updated the entire model for 100 epochs. We used the AdamW optimizer with a learning rate of 1e-4 and the batch size was 16 throughout the entire training. The resolutions of generated images were $256 \times 256$. The sampler was DDIM with 50 steps.
\\

\noindent \textbf{Uncertainty-aware model architecture.} 
In our experiments, the specific model is selected for different medical image analysis tasks and provides the model uncertainty estimation for predictions \cite{aminiDeep2020,zou2022tbrats}. The model uncertainty usually exhibits a high value when processing handheld device data.
We used the stacked hourglass network \cite{newellStacked2016} (SHN) for the detection task. Our SHN only stacked two single hourglasses to avoid overfitting and downsampled the feature map four times in each hourglass to extract high-level spatial context. We used a basic U-Net \cite{ronnebergerUNet2015} for the segmentation task. In particular, we incorporated an additional branch into both SHN and U-Net to estimate model uncertainty, which can be simply inserted into the head module without altering the fundamental structure of models. The uncertainty can be used to measure the distance of image feature distributions between handheld and standard device data.
\\

\noindent \textbf{Train uncertainty-aware model using source data.} 
Given the source data $D_s=\{x_s^i,y_s^i\}_{i=1}^N$ and the model  $f_\theta(\cdot)$, the learning objective is specific to the task. We used the MSE loss for the detection task
\begin{equation}
\label{det_l}
    L_{task}=\frac{1}{N}\sum_{i=1}^N \big\|y_s^i-f_\theta(x_s^i)\big\|^2
\end{equation}
where the annotation $y_s$ is heatmaps. For the segmentation task, we used the cross-entropy loss and dice loss to update the U-Net
\begin{equation}
\label{seg_l}
    L_{task} = \sum_{n=1}^C-y_s^n\log(\hat{y}) + 1-\frac{2*y_s*\hat{y}+1}{y_s+\hat{y}+1}
\end{equation}
where $\hat{y}=f_\theta(x_s)$ is the predicted mask.

Furthermore, we not only required accurate predictions but also demanded the uncertainty of the predictions. We pushed the model to directly estimate model uncertainty by predicting the parameters of the Normal-Inverse-Gamma (NIG) distribution \cite{aminiDeep2020}, which is a conjugate prior distribution on the source domain distribution. We considered $y_s\sim\mathcal{N}(\mu,\sigma^2)$, $\mu\sim\mathcal{N}(\gamma,\sigma^2\omega^{-1})$ and $\sigma^2\sim\Gamma^{-1}(\alpha, \beta)$, where the $\Gamma^{-1}$ is the gamma function. The distribution of $y_s$ follows the form of an $NIG(\gamma, \omega, \alpha, \beta)$ distribution
\begin{equation}
\label{NIG}
    p(\mu, \sigma^2|\gamma,\omega,\alpha,\beta)=\frac{\beta^{\alpha}\sqrt{\omega}}{\Gamma(\alpha)\sqrt{2\pi\sigma^2}}\left(\frac{1}{\sigma^2}\right)^{\alpha+1}\exp\left\{-\frac{2\beta+\omega(\gamma-\mu)^2}{2\sigma^2}\right\}
\end{equation}
where $\gamma\in \mathbb{R}$, $\omega>1$ and $\beta>0$. During the training phase, the following negative log-likelihood loss was applied to address the NIG distribution
\begin{equation}
\label{NIG_l}
    L_{NIG}=\frac{1}{2}\log\left(\frac{\pi}{\omega}\right)-\alpha\log(\Omega)+(\alpha+\frac{1}{2})\log((y-\gamma)^2\omega+\Omega)+\log(\Theta)
\end{equation}
where $\Omega=2\beta(1+\omega)$ and $\Theta=\left(\frac{\Gamma(\alpha)}{\Gamma(\alpha+\frac{1}{2})}\right)$. To penalize the incorrect evidence, an regularizer $L_R=|y_i-\gamma|\cdot(2\omega+\alpha)$ was introduced into the total loss
\begin{equation}
\label{un_l}
    L_{un}=L_{NIG}+\lambda L_{R}
\end{equation}
where coefficient $\lambda$ trades off these two loss terms. Finally, we replaced the deterministic output of the model with a NIG distribution  
\begin{equation}
\label{NIG_model}
    f_\theta(x_s)=NIG(\gamma, \omega, \alpha, \beta)
\end{equation}
where the prediction $\hat{y}=\mathbb{E}(u)=\gamma$ and the model uncertainty map is defined as
\begin{equation}
\label{un_map}
    U =\frac{\beta}{\omega(\alpha-1)}
\end{equation}
we computed the model uncertainty as:
\begin{equation}
\label{un_model_det}
    U^{model}_{det}=U_{\hat{y}}, \ \hat{y}\ is\ coordinate
\end{equation}
\begin{equation}
\label{un_model_seg}
    U^{model}_{seg}=\frac{\sum_{\hat{y}==1}U}{\sum_{\hat{y}==1}}, \ \hat{y}\ is\ mask
\end{equation}
\\

\noindent \textbf{Produce condition images.}
The condition image provides spatial context and guides the generation of the diffusion model. We used the binary masks for the spine ultrasound data using a global threshold, and the noise perturbation for the carotid ultrasound data by adding random Gaussian noise. Both implementations do not involve any learnable parameters and can produce different condition images by modifying the thresholds. 
For the diffusion model trained on source data, we only produced an optimal condition image paired with a source image using annotations, which can provide the most precise and concise spatial context. During the alignment phase, we produced multiple condition images within a suitable threshold range. The goal of the alignment phase is to find the best condition image from these options, which can guide the generative model to achieve the optimal alignment for the target image.
\\

\noindent \textbf{Align target images to adapt models trained on source data during inference.}
Our adaptation scheme consists of two phases: style alignment and selection. The alignment phase aims to align the target image to match the source domain style while preserving the original structure. To achieve this, $m$ condition images with different spatial contexts were produced for a target image to control the generative model, with the key challenge being to find the optimal condition image that was simultaneously aligned in style and preserved its structure. Hence, during the selection phase, we separately utilized the model and prediction uncertainties to quantitatively assess style alignment and structure preservation. 

The model uncertainty represents the model's confidence in its predictions and usually exhibits higher values for target domain data. Our uncertainty-aware model incorporates the uncertainty estimation module and can provide a certain uncertainty value for each prediction.  
The prediction uncertainty represents the stability among multiple outcomes. We utilized the same condition image to control the generative model synthesizing $n$ source-like images; the spatial structures of the $n$ generated images should be consistent if this condition image provides a precise spatial context. In other words, the uncertainty-aware model should output consistent predictions for these $n$ images. Given the predictions $Y=\{\hat{y_1},\hat{y_2},\dots,\hat{y_{n}}\}$ of $n$ generated image, the prediction uncertainty is defined as:
\begin{equation}
\label{un_pre_det}
    U^{pre}_{det}=\frac{1}{n}\sum^n_{i=1}\|\hat{y_i}-\bar{y}\|,\ \bar{y}=\frac{1}{n}\sum^n_{i=1}\hat{y_i} 
\end{equation}
\begin{equation}
\label{un_pre_seg}
    U^{pre}_{seg}=1-\frac{\hat{y_1}\cap \hat{y_2}\dots \cap \hat{y_n}}{\hat{y_1}\cup \hat{y_2}... \cup \hat{y_n}}
\end{equation}

Following the scheme, we sequentially generated $m$ condition images, $m*n$ aligned images, $m*n$ predictions, model uncertainty matrix $U^{model}_{m\times n}$, and prediction uncertainty vector $U^{pre}_{m\times 1}$ for a target image. The model uncertainty matrix was transformed into a vector $U_{m\times 1}^{model}$ by taking the average. Finally, we utilized two uncertainty vectors to assess the generation quality of each condition image. We first filtered out the condition images with uncertainty values above thresholds $T^{model}$ and $ T^{pre}$, and selected the optimal condition image with the lowest prediction uncertainty from the remaining images. The $T^{model}$ was computed based on the uncertainty of the model's predictions on the source data and found to be 0.0099 and 0.1200 for spine and carotid ultrasound images, respectively. The $T^{pre}$ was found to be 5 and 0.3. Once the optimal condition image and its $n$ predictions were identified, we selected the one that had minimal model uncertainty as the final prediction.
\\

\noindent \textbf{Model evaluation.} 
To evaluate TISA performance on the detection task, we considered the predicted landmarks as positive if they were within 10 pixels of the ground truth \cite{newellStacked2016}, and then computed the accuracy, recall, and F1 score. The mean error was the average distance between all predicted landmarks and ground truths. For the segmentation task, we used the Dice score and IoU for evaluation. To evaluate TISA's ability for automatic diagnosis, we did correlation and Bland–Altman analysis \cite{giavarina2015understanding} for both spinal curvature and CMIT measurements.

\bibliography{references}

\section*{Data availability}
Carotid data from CUBS are available at (\href{https://data.mendeley.com/datasets/fpv535fss7/1}{https://data.mendeley.com/datasets/fpv535fss7/1}). Individual-level patient data can be accessible with the consent of the data management committee from institutions and are not publicly available. Requests for the non-profit use of ultrasound images and related clinical information should be sent to R.Z.

\section*{Code availability}
All codes are available at \href{https://github.com/zenghy96/TISA}{https://github.com/zenghy96/TISA}.

\section*{Acknowledgements}
The authors would like to extend sincere gratitude to Glenrose Hospital and Tongren Hospital for providing the subjects’ ultrasound scans in this study. And the authors also are profusely grateful for the sponsorship from the Natural Science Foundation of China (NSFC), Grant No. 12074258.

\section*{Author contributions statement}
H.Z.: conceptualization, methodology, data collection, experimental deployment, software, writing—original draft. 
K.Z. and Z.C: methodology, experimental deployment, writing—review and editing. 
Y.C., H.C, and H.Z: data collection and annotation and curation. 
K.Z., W.M., R.S.M.G, and Y. L: writing—review and editing.
C.J.: clinical assessment and annotation and curation, clinical support, writing-review and editing. 
R.Z. and H.F.: supervision, project administration, methodology, writing—review and editing.

\section*{Additional information}
Supplementary information is provided.



\end{document}